\documentclass[11pt,twoside]{article}
\usepackage{epsfig}
\usepackage{graphicx}
\begin{document}
\newcommand{\M}{\mbox{m}}
\newcommand{\n}{\mbox{$n_f$}}
\newcommand{\EE}{\mbox{ee}}
\newcommand{\EP}{\mbox{e$^+$}}
\newcommand{\EM}{\mbox{e$^-$}}
\newcommand{\EPEM}{\mbox{e$^+$e$^-$}}
\newcommand{\EMEM}{\mbox{e$^-$e$^-$}}
\newcommand{\GG}{\mbox{$\gamma\gamma$}}
\newcommand{\GE}{\mbox{$\gamma$e}}
\newcommand{\GP}{\mbox{$\gamma$e$^+$}}
\newcommand{\TEV}{\mbox{TeV}}
\newcommand{\GEV}{\mbox{GeV}}
\newcommand{\LGG}{\mbox{$L_{\gamma\gamma}$}}
\newcommand{\LGE}{\mbox{$L_{\gamma e}$}}
\newcommand{\LEE}{\mbox{$L_{ee}$}}
\newcommand{\WGG}{\mbox{$W_{\gamma\gamma}$}}
\newcommand{\WGE}{\mbox{$W_{\gamma e}$}}
\newcommand{\EV}{\mbox{eV}}
\newcommand{\CM}{\mbox{cm}}
\newcommand{\MM}{\mbox{mm}}
\newcommand{\NM}{\mbox{nm}}
\newcommand{\MKM}{\mbox{$\mu$m}}
\newcommand{\SEC}{\mbox{s}}
\newcommand{\CMS}{\mbox{cm$^{-2}$s$^{-1}$}}
\newcommand{\MRAD}{\mbox{mrad}}
\newcommand{\IND}{\hspace*{\parindent}}
\newcommand{\E}{\mbox{$\epsilon$}}
\newcommand{\EN}{\mbox{$\epsilon_n$}}
\newcommand{\EI}{\mbox{$\epsilon_i$}}
\newcommand{\ENI}{\mbox{$\epsilon_{ni}$}}
\newcommand{\ENX}{\mbox{$\epsilon_{nx}$}}
\newcommand{\ENY}{\mbox{$\epsilon_{ny}$}}
\newcommand{\EX}{\mbox{$\epsilon_x$}}
\newcommand{\EY}{\mbox{$\epsilon_y$}}
\newcommand{\BI}{\mbox{$\beta_i$}}
\newcommand{\BX}{\mbox{$\beta_x$}}
\newcommand{\BY}{\mbox{$\beta_y$}}
\newcommand{\SX}{\mbox{$\sigma_x$}}
\newcommand{\SY}{\mbox{$\sigma_y$}}
\newcommand{\SZ}{\mbox{$\sigma_z$}}
\newcommand{\SI}{\mbox{$\sigma_i$}}
\newcommand{\SIP}{\mbox{$\sigma_i^{\prime}$}}
\newcommand{\lsim}{\raisebox{-0.07cm}{$\,
\newcommand{\NC}{\mbox{${\cal NC}$}}
\newcommand{\CC}{\mbox{${\cal CC}$}}
\stackrel{<}{{\scriptstyle\sim}}\, $}}
\newcommand{\gsim}{\raisebox{-0.07cm}{$\,
\stackrel{>}{{\scriptstyle\sim}}\, $}}
\def\b{\beta}
\def\g{\gamma}
\def\SM{$\mathcal{SM}$}
\def\MSSM{$\mathcal{MSSM}$}
\def\2HDM{$2\mathcal{HDM}$}
\def\h{\rm h}
\def\ccbar{\overline{\mbox c}\mbox{c}}
\def\bbbar{\overline{\mbox b}\mbox{b}}
\def\qqbar{\overline{\mbox q}\mbox{q}}
\def\ccbarg{\overline{\mbox c}\mbox{cg}}
\def\bbbarg{\overline{\mbox b}\mbox{bg}}
\def\BR{\rm BR}
\newcommand{\be}{\begin{equation}}
\newcommand{\ee}{\end{equation}}
\newcommand{\bea}{\begin{eqnarray}}
\newcommand{\eea}{\end{eqnarray}}
\newcommand{\bear}{\begin{equation}\begin{array}}
\newcommand{\eear[1]}{\end{array}{#1}\end{equation}}
\newcommand{\dst}{\displaystyle}
\newcommand{\fordef}{\stackrel{def}{=}}
\newcommand{\bm}{\boldmath}
\newcommand{\fr}[2]{\frac{{\displaystyle #1}}{{\displaystyle #2}}}
\newcommand{\nn}{\nonumber}
\newcommand{\pa}{\partial}
\newcommand{\la}{\langle}
\newcommand{\ra}{\rangle}
\newcommand{\fn}[1]{\footnote{{\normalsize #1}}}
\def\emline#1#2#3#4#5#6{%
       \put(#1,#2){\special{em:moveto}}%
       \put(#4,#5){\special{em:lineto}}}

\newenvironment{Itemize}{\begin{list}{$\bullet$}%
{\setlength{\topsep}{0.2mm}\setlength{\partopsep}{0.2mm}%
\setlength{\itemsep}{0.2mm}\setlength{\parsep}{0.2mm}}}%
{\end{list}}
\newcounter{enumct}
\newenvironment{Enumerate}{\begin{list}{\arabic{enumct}.}%
{\usecounter{enumct}\setlength{\topsep}{0.2mm}%
\setlength{\partopsep}{0.2mm}\setlength{\itemsep}{0.2mm}%
\setlength{\parsep}{0.2mm}}}{\end{list}}
\newcommand{\sw}{\mbox{$\sin\Theta_W\,$}}
\newcommand{\cw}{\mbox{$\cos\Theta_W\,$}}
\newcommand{\epe}{\mbox{$e^+e^-\,$}}
\newcommand{\ggam}{\mbox{$\gamma\gamma\,$}}
\newcommand{\egam}{\mbox{$e\gamma\,$}}
\newcommand{\gewnu}{\mbox{$e\gamma\to W\nu\,$}}
\newcommand{\eeww}{\mbox{$e^+e^-\to W^+W^-\,$}}
\newcommand{\ggww}{\mbox{$\gamma\gamma\to W^+W^-\,$}}
\newcommand{\ggzz}{\mbox{$\gamma\gamma\to ZZ\,$}}
\newcommand{\egeh}{\mbox{$e\gamma\to eH\,$}}
\newcommand{\geeww}{\mbox{$e\gamma\to e W^+W^-\,$}}
\newcommand{\beq}{\begin{equation}}
\newcommand{\eeq}{\end{equation}}
\newcommand{\beqn}{\begin{eqnarray}}
\newcommand{\eeqn}{\end{eqnarray}}
\newcommand{\lum}[1]{{\rm luminosity} $ #1 $ cm$^{-2}$ s$^{-1}\,$}
\newcommand{\intlum}[1]{{\rm annual luminosity} $ #1$  fb$^{-1}\,$}
\newcommand{\mw}{\mbox{$M_W\,$}}
\newcommand{\mww}{\mbox{$M_W^2\,$}}
\newcommand{\mh}{\mbox{$M_H\,$}}
\newcommand{\mhh}{\mbox{$M_H^2\,$}}
\newcommand{\mz}{\mbox{$M_Z\,$}}
\newcommand{\mzz}{\mbox{$M_Z^2\,$}}
\newcommand{\sigmaw}{\mbox{$\sigma_W\,$}}
\newcommand{\sww}{\mbox{$\sin^2\Theta_W\,$}}
\newcommand{\cww}{\mbox{$\cos^2\Theta_W\,$}}
\newcommand{\ptr}{\mbox{$p_{\bot}\,$}}
\newcommand{\ptrs}{\mbox{$p_{\bot}^2\,$}}
\newcommand{\lgam}{\mbox{$\lambda_{\gamma}$}}
\newcommand{\lga}[1]{\mbox{$\lambda_{#1}$}}
\newcommand{\lggam}[2]{\mbox{$\lambda_{#1}\lambda_{#2}$}}
\newcommand{\lgg}{\lambda_1\lambda_2}
\newcommand{\lel}{\mbox{$\lambda_e$}}
\newcommand{\ggh}{\mbox{$\gamma\gamma\to hadrons$}}
\newcommand{\pair}[1]{\mbox{$#1 \bar{#1}$}}
\newcommand{\ZZ}{\mbox{ZZ}} \newcommand{\WW}{\mbox{WW}}
\newcommand{\Z}{\mbox{Z}} \newcommand{\W}{\mbox{W}}
\setcounter{page}{1}
\title{\vspace{-2.5cm} \hspace*{9cm} 
{\large BUDKER-INP 2003-7} \\[7mm]
MEASUREMENT OF \GG\ and \GE\ LUMINOSITIES
  AT PHOTON COLLIDERS~\thanks{Talk at the Intern. Workshop on Physics
    and Detectors at Linear Collider, LCWS2002, August 26{-}30, 2002, Jeju
Island, Korea.}} 
\author{A.V.Pak$^{1)}$, D.V.Pavluchenko$^{2)}$,
  S.S.Petrosyan$^{2)}$,
  V.G.Serbo$^{1)}$,  V.I.Telnov
\thanks{Corresponding author: e-mail address: telnov@inp.nsk.su, 
telnov@mail.desy.de}~ $^{2)}$ \\ \\
  {\it $^{1)}$Novosibirsk State University, 630090, Novosibirsk, Russia}\\
  {\it $^{2)}$Budker Institute of Nuclear Physics, 630090,
    Novosibirsk, Russia}} \date{} \maketitle \vspace{-1.3cm}
\begin{abstract} \vspace{-0.3cm}
   Methods of  \GG, \GE\ luminosities measurement at photon
   colliders based on Compton scattering of laser photons on high energy
   electrons at linear colliders are considered.
\end{abstract}
%
\vspace{-0.4cm}
\section{Introduction}
\vspace{-0.4cm}
In addition to \EPEM\ physics, linear colliders provide a unique
opportunity to study \GG\ and \GE\ interactions at high energies and
luminosities~\cite{GKST83,GKST84}. High energy photons can be
obtained using Compton backscattering of laser light off high energy
electrons. This option is included in Technical Design of the linear
collider TESLA~\cite{TESLATDR} and is considered for all other project of
linear colliders~\cite{NLC,JLC,CLIC}.

\vspace{-3.5mm} 

Spectrum of photons after Compton scattering is broad with
characteristic peak at maximum energies. Photons can have circular or
linear polarizations depending on their energies and polarizations of
initial electrons and laser photons. Due to angle-energy correlation,
in Compton scattering the \GG\ luminosity can not be described by
convolution of some photon spectra.  Due to complexity of processes in
the conversion and interaction regions an accuracy of prediction by
simulation will be rather poor, therefore one should measure all
luminosity properties experimentally.

\vspace{-3.5mm} 

Below we consider some general  features of \GG, \GE\ 
processes, give example of  luminosity spectra
and then consider methods for measurement of \GG, \GE\  luminosities. 
\vspace{-0.5cm}
\section{General features of luminosities and cross sections}
\vspace{-0.5cm}
In general case the number of events in \GG\ collision is given 
by~\cite{GKST84}
\begin{equation}
\vspace{-0.3cm}
d\dot{N}_{\GG\ \to X} = dL_{\GG} \sum_{i,j=0}^3 \langle \xi_i \tilde{\xi_j}
\rangle \sigma_{ij}, 
\vspace*{-0.1cm}
\label{dn1}
\end{equation}
where $\xi_i$ are Stokes parameters, $\xi_2\equiv \lambda_{\gamma}$ is
the circular polarization, $\sqrt{\xi_1^2+\xi_3^2} \equiv l_{\gamma}$  
the linear
polarization and $\xi_0 \equiv 1$.  Since photons have wide
spectra and various polarizations, in general case one has to measure
16 two dimensional luminosity distributions $d{\,^2}L_{ij}/d\omega_1
d\omega_2$,  $dL_{ij} = dL_{\GG} \langle \xi_i \tilde{\xi_j} \rangle$,
where the tilde sign marks the second colliding beam. 

\vspace{-2mm}

Among 16 cross sections $\sigma_{ij}$ there are three most important
which do not vanish after averaging over spin states of final
particles and azimuthal angles, that are~\cite{GKST84} \vspace{-0.1cm}
$$\sigma^{np} \equiv \sigma_{00} =
\frac{1}{2}(\sigma_{\parallel}+\sigma_{\perp})= 
\frac{1}{2}(\sigma_0+\sigma_2) $$ 
\vspace{-0.3cm}
\begin{equation} \tau^c \equiv \sigma_{22} =
  \frac{1}{2}(\sigma_0-\sigma_2) \;\;\;\;
\tau^l \equiv   \frac{1}{2}(\sigma_{33} - \sigma_{11}) = 
\frac{1}{2}(\sigma_{\parallel}-\sigma_{\perp}) 
\end{equation}
\vspace{-0.0cm}
Here $\sigma_{\parallel},\sigma_{\perp}$ are cross sections for
collisions of linearly polarized photons with parallel and orthoganal
relative polarizations and  $\sigma_0$ and $\sigma_2$ are cross sections
for collisions of photons with  $J_z$ of two photons equal 0 and
2, respectively.

\vspace{-2mm}

If only these three cross sections are of interest than (\ref{dn1}) can
be written as 
\begin{equation}
\vspace{-0.1cm}
d\dot{N}_{\GG\ \to X} = dL_{\GG}\; (d\sigma^{np} + \langle \xi_2 \tilde{\xi_2}
\rangle d\tau^c + \langle \xi_3 \tilde{\xi_3} - 
\xi_1 \tilde{\xi_1} \rangle d \tau^l)\,.
\vspace{-0.1cm}
\end{equation}

\vspace{-2mm}

Substituting $\xi_2 \equiv \lambda_{\gamma}$, $ \tilde{\xi_2} \equiv
\tilde{\lambda}_{\gamma}$, $\xi_1 \equiv l_{\gamma} \sin 2\gamma$ ,
$\tilde{\xi_1} \equiv -\tilde{l}_{\gamma} \sin 2\tilde{\gamma}$,
$\xi_3 \equiv l_{\gamma} \cos 2\gamma$, $\tilde{\xi_3} \equiv
\tilde{l}_{\gamma} \cos 2 \tilde{\gamma}$ and 
$\Delta\phi=\gamma -\tilde{\gamma}$ (azimuthal angles for linear
polarizations are defined relative
to one $x$ axis), we get
\begin{eqnarray}
d\dot{N} & = & dL_{\GG}( d\sigma^{np} + \lambda_{\gamma} \tilde{\lambda}_{\gamma}\;
d\tau^c +   l_{\gamma} \tilde{l}_{\gamma} \cos{2\Delta\phi} \;d\tau^l)
 \nonumber \\ 
&\equiv  & dL_{\GG}\;  d\sigma^{np} + (dL_0 - dL_2) d\tau^c + (dL_{\parallel}
 -dL_{\perp})\,d\tau^l   \nonumber \\ 
& \equiv  & dL_0  d\sigma_0 + dL_2  d\sigma_2 + (dL_{\parallel} -dL_{\perp})\,
d\tau^l   \nonumber \\ 
&\equiv  & dL_{\parallel}\,  d\sigma_{\parallel} + dL_{\perp}\,  d\sigma_{\perp}
 + (dL_0 - dL_2)\, d\tau^c\,,
\label{dn}
\end{eqnarray}
where \,
$dL_0 = dL_{\gamma}(1+\lambda_{\gamma} \tilde{\lambda}_{\gamma})/2\;,
dL_2 = dL_{\gamma}(1-\lambda_{\gamma} \tilde{\lambda}_{\gamma})/2\;,
dL_{\parallel} = dL_{\gamma} (1+ l_{\gamma}
\tilde{l}_{\gamma}\cos{2\Delta\phi})/2\;$, 
$dL_{\perp} = dL_{\gamma}(1- l_{\gamma} \tilde{l}_{\gamma}\cos{2\Delta\phi})/2\,.$

\vspace{-2mm}

So, one should measure (not only in a general case) $dL_{\GG}$, $\langle
\lambda_{\gamma} \tilde{\lambda}_{\gamma}\rangle$, $\langle l_{\gamma}
\tilde{l}_{\gamma}\rangle $ or alternatively $dL_0, dL_2, dL_{\parallel},
dL_{\perp}$. If both photon beams have no linear polarization or no
circular polarization, the luminosity 
can be decomposed in two parts: $L_0$ and $L_2$, or
$L_{\parallel}$ and $L_{\perp}$, respectively.

\vspace{-2mm}

For example, for scalar resonances such as the Higgs boson, 
$\sigma_2=0$, while 
$\sigma_{\parallel} = \sigma_0$, $\sigma_{\perp} = 0$ for $CP=1$ and  
$\sigma_{\perp} = \sigma_0$, $\sigma_{\parallel} = 0$ for $CP=-1,$
then \\[-3mm]
\begin{equation}
\vspace{-0.2cm}
d\dot{N} = dL_{\GG} \,d\sigma^{np} (1+\lambda_{\gamma}
\tilde{\lambda}_{\gamma} \pm l_{\gamma} \tilde{l}_{\gamma}\cos{2\Delta\phi})\,.
\vspace{-0.3cm}
\end{equation}

In \GE\ collisions general formulae are similar to (\ref{dn1}) but
Stokes parameters for the second particle are replaced by the electron
spin vector {\boldmath{$\zeta$}} (the double mean electron helicity
$2\lambda_e=\zeta_3)$. Besides $dL_{\GE}$ the most important
characteristic is $\langle \lambda_e \lambda_{\gamma} \rangle$ (as
function of $E_e$ and $E_{\gamma}$), but in some cases $\lambda_e$ and
$\lambda_{\gamma}$ should be known separately, in the process $\GE\to
W \nu$, for instance.  

\vspace{-2mm}

\begin{figure}[!hbt]
\centering
\vspace*{-0.8cm}
\hspace*{-0.0cm} \epsfig{file=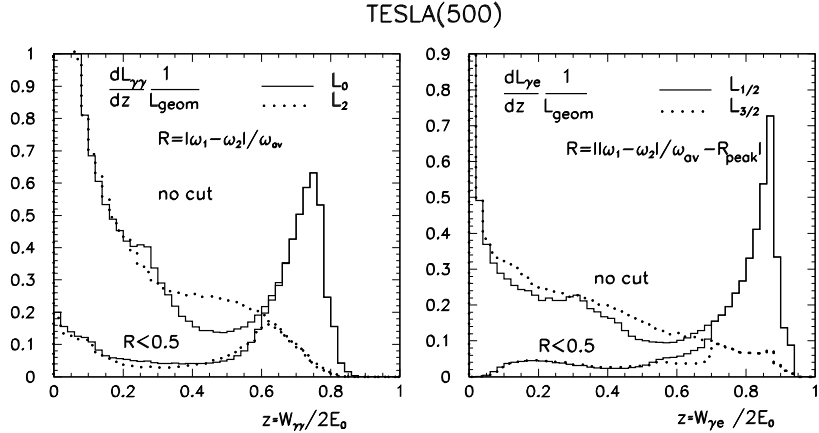,width=13.5cm,angle=0}
\vspace{-1.4cm}
\caption{Left: \GG\ luminosity spectra; right: \GE\ luminosity spectra
  at TESLA. 
Solid lines for  $J_z$ of two colliding photons equal to 0,   dotted lines
  for  $J_z=2$ (1/2 and 3/2 in case of \GE\ collisions).}
\label{Ldist250}
\vspace*{-0.9cm}
\end{figure}

Expected \GG, \GE\ luminosity spectra at TESLA(500) are
presented in Fig.\ref{Ldist250}~\cite{TEL2000,TESLATDR,TEL2002}. The
 luminosities in the high energy peaks are about $10^{34}\,\CMS$
 both for \GG\ and \GE\ collisions~\cite{TESLATDR}.
\vspace{-0.2cm}
\section{Measurement of \GG\ luminosities}
\vspace{-0.5cm} The best process for the measurement of the \GG\ 
luminosity is $\GG\to l^+l^-$ ($l=\mbox{e},\mu$)
\cite{GKST84,TEL93,TEL95,TESLATDR}. Another QED process with large
cross section is $\GG\to l^+l^- l^+l^-$~\cite{GKST83,Kapusta}.  The
process $\GG\ \to W^+W^-$ was also suggested for the luminosity
measurement~\cite{YASUI}, but at first its cross section has to be
measured as it may differ from prediction of the Standard Model and
presents special physics interest by itself.

\vspace{-0.2cm}

{\boldmath \underline{$ \GG\ \to l^+l^-$}.}  The cross section in
c.m.s. at not too small angles (see \cite{TESLATDR})
\begin{equation}
d\sigma = \frac{2\pi\alpha^2}{\WGG^2}\left[(1-\lambda_{\gamma}\tilde{\lambda}_{\gamma})
\frac{(1+\cos^2 \theta)}{(1-\cos^2 \theta)}
-l_{\gamma}\,\tilde{l}_{\gamma} 
\cos{(4 \varphi - 2 (\gamma +\tilde{\gamma}))}\right]\frac{d\varphi}{2\pi} d(\cos \theta)\,,
\end{equation}
where $\lambda_{\gamma},\tilde{\lambda}_{\gamma}$ the circular
polarizations, $l_{\gamma},\tilde{l}_{\gamma}$ linear polarizations of
photons,  $\gamma,\tilde{\gamma}$ azimuthal angles
  of photon linear polarizations,  $\varphi$  the azimuthal angle of the lepton plane.
One can see that the cross section is zero for
$\lambda_{\gamma}=\tilde{\lambda}_{\gamma}= \pm 1$ (or $J_z=0$). Thus, lepton pairs are
produced only in collisions of photons with total $J_z=2$.  The
suppression factor for $J_z=0$ is $m_l^2/W^2_{\GG}$~\cite{TESLATDR}.
This process can measure only $L_2$, for measurement of $L_0$ one can
change the polarization of one beam to opposite for part of the
time~\cite{TEL93,TEL95}. The product of linear polarizations $\langle
l_{\gamma}\tilde{l}_{\gamma} \rangle$ can be measured by azimuthal variation of the cross
section at the large angles $\theta$.  The total cross section
\vspace{-0.2cm}
\begin{equation}
\sigma_2 (|\cos \theta| < a) \approx \frac{4\pi\alpha^2}{W^2_{\gamma\gamma}}
\left[2 \ln \left(\frac{1+a}{1-a}\right) -2a \right].
\vspace{-0.cm}
\end{equation}
 For example, $\sigma(|\cos\theta|<0.9)\approx
 10^{-36}/W^2_{\GG}[\mathrm{TeV}]\; \CM^2$,
and at TESLA(500) for $10^7$ sec one can expect about $10^6$ of
 \EPEM\ and $\mu^+\mu^-$ events  in the high energy peak. 

\vspace{-0.2cm}

\begin{figure}[!hbt]
\centering
\vspace*{-0.5cm}
\hspace*{-0.5cm} \epsfig{file=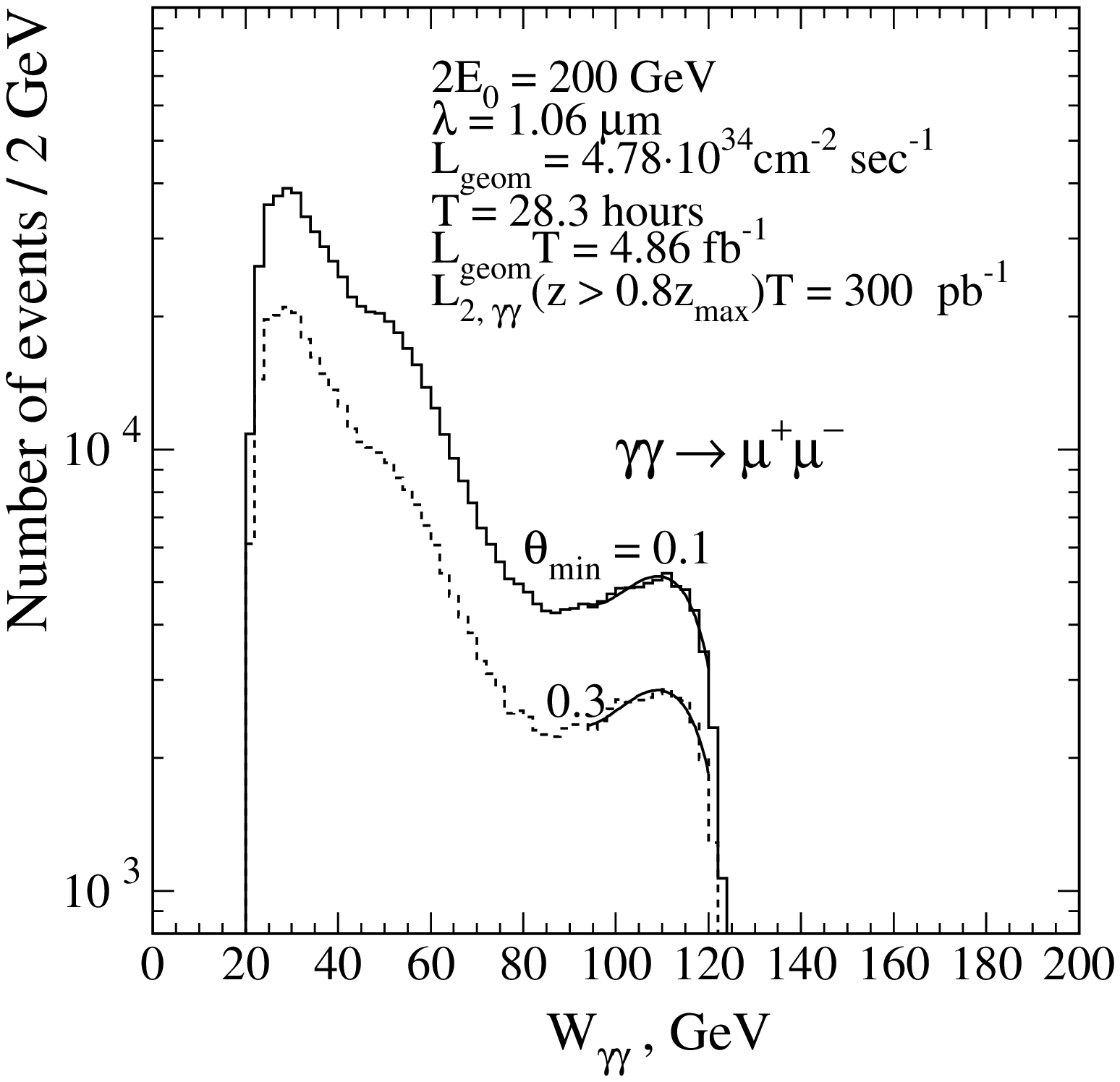,width=5.8cm,angle=0} \hspace*{-0.0cm}
 \epsfig{file=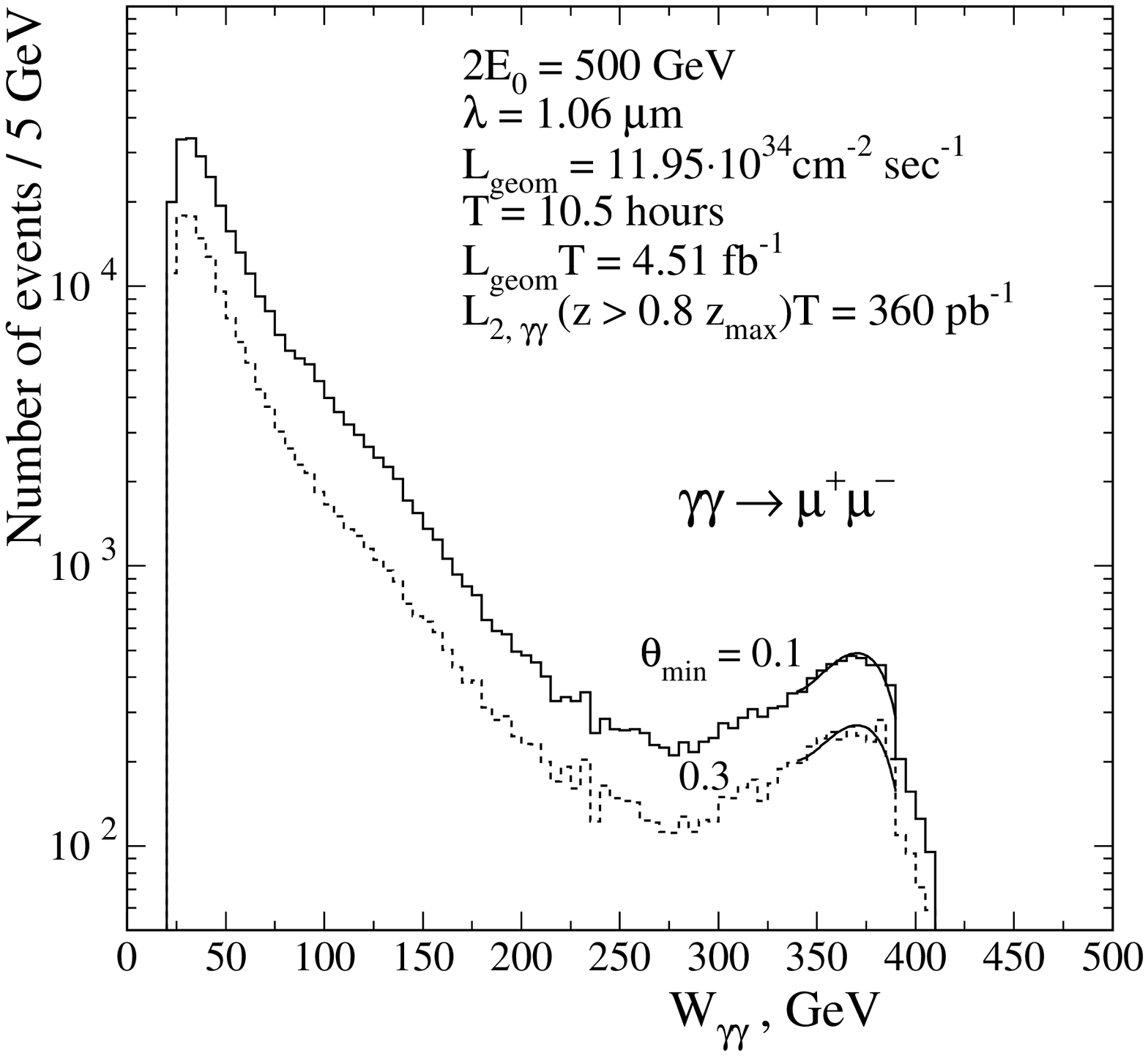,width=6cm,angle=0}
\vspace{-0.7cm}
\caption{Distribution of pairs on the invariant mass in the process $\GG\
  \to \mu^+\mu^-$; left: $2E_0=200$ GeV, right: $2E_0=500$ GeV.  }
\vspace{-0.2cm}
\label{mumu}
\end{figure}

The invariant mass spectrum of $\mu^+\mu^-$ pairs in the detector is
shown in Fig.\ref{mumu}. 
For the minimum detection angle $\theta = 0.3$ rad and $10^7$ sec run
the expected statistical accuracy of the peak value of $dL/dW_{\GG}$ is
about 0.07 \% and 0.14 \% for $2E_0=200$ and 500 GeV, respectively.
This is much better than necessary for the Higgs study where the
accuracy for branchings not better than 1\% is expected.

 \vspace{-0.2cm}
 
\begin{figure}[!hbt]
\vspace{-0.7cm}
\hspace*{0cm}\begin{minipage}[b]{0.45\linewidth}
\centering
\vspace*{-0.0cm} 
\hspace*{.0cm} \epsfig{file=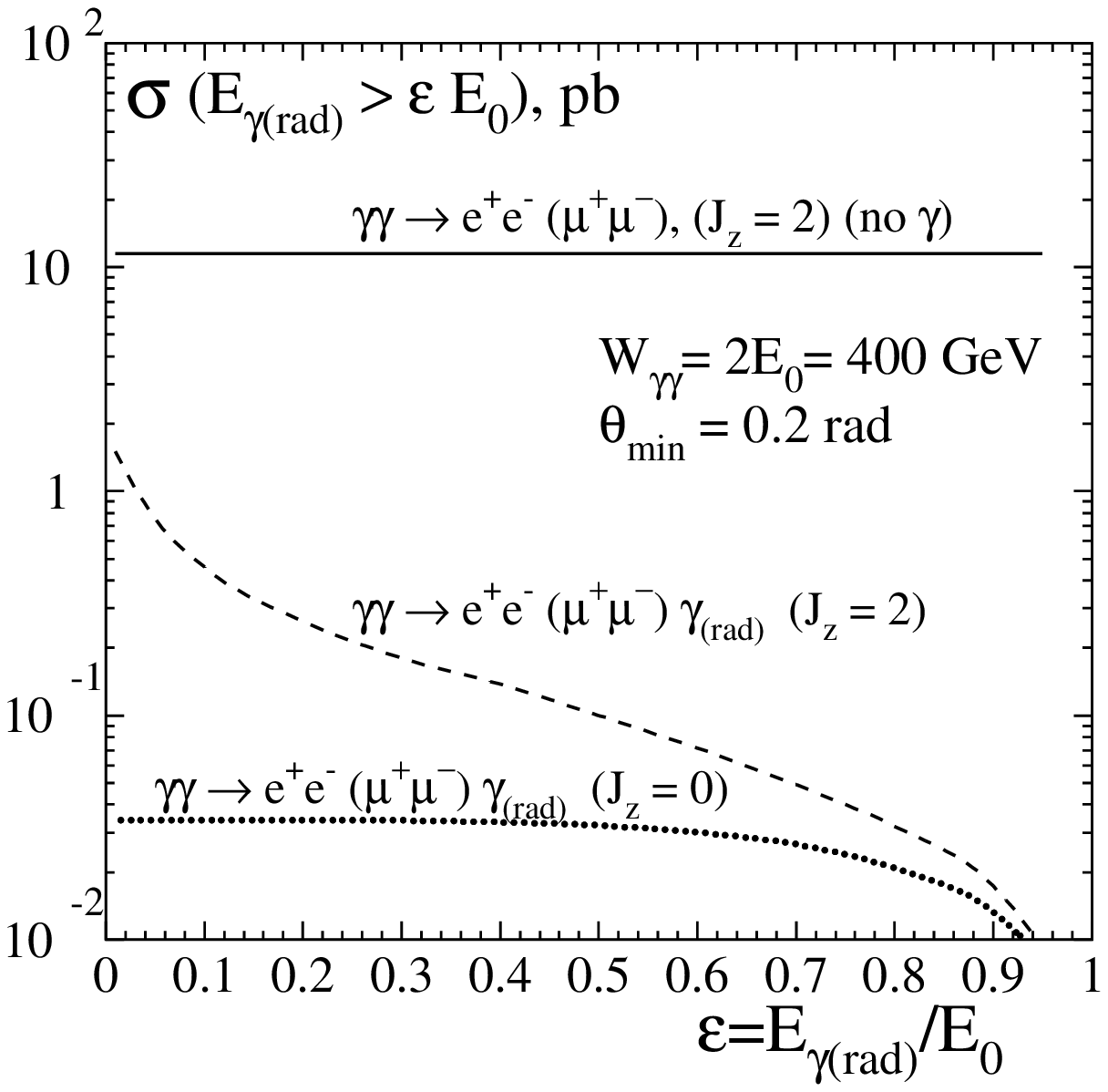,height=6.cm,angle=0} 
\vspace*{-0.8cm}
\caption{ \label{gge} The cross section of the process $\GG\to l^+l^-\gamma$ vs
  the minimum energy of the emitted photons.   }
\vspace*{0.0cm}
\end{minipage}%
\hspace*{0.5cm} \begin{minipage}[b]{0.45\linewidth}
\centering
\hspace*{-0.5cm} \epsfig{file=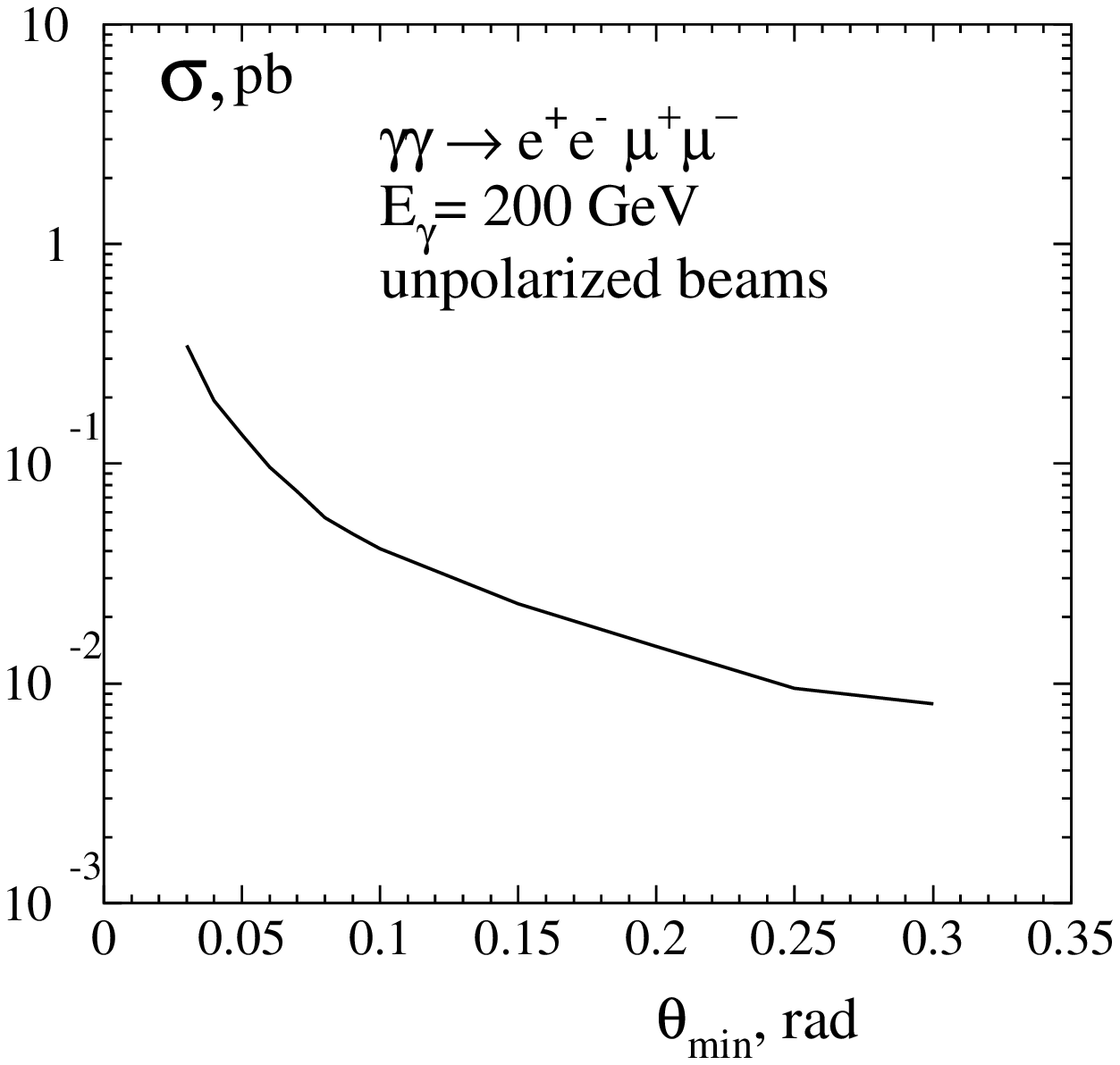,width=6.cm,angle=0}
\vspace*{-0.8cm}
\caption{\label{eemm} 
 The cross section of the process $\GG\to e^+e^-\mu^+\mu^-$ vs
  the minimum polar angle of all final particles.} 
 \vspace*{-0.cm}
\end{minipage}
\end{figure}

 {\boldmath \underline{$\GG\ \to l^+l^-\gamma$}}.  This process is of
 interest because the photon emission removes the suppression of the
 lepton production in the case of $J_z=0$. The cross section for this
 process as a function of the minimum photon energy obtained by the
 ComHEP code~\cite{comhep} is presented in Fig.\ref{gge}. Indeed, the
 cross section for this case is not negligible, but  lower
 than for $J_z=2$ without photons by a factor of 300.

\vspace{-5mm}
{\boldmath \underline{$\GG\ \to l^+l^-l^+l^- (l=e,\mu)$}}. This
process is of interest~\cite{GKST83,Kapusta} because it has large total cross
section and weakly depends on photon polarizations, therefore it could
be used together with $\GG\to l^+l^-$ for measurement of $L_0$ and
$L_2$ simultaneously. Unfortunately, at large angles, where particle
momenta can be measured with a high accuracy, the cross section is
rather small. The cross section obtained by ComHEP for the process
$\GG\ \to e^+e^-\mu^+\mu^-$ at the photon energies $2\times 200$ GeV
is shown in Fig.~\ref{eemm}.  At small angles main contribution gives
the  peripheral diagram and the cross section depends on the minimum
angle as $1/\theta^2$.  At the angles above 0.1 rad main contribution
give  diagrams where the second  pair is emitted by one of
leptons of the first pair, here one can expect the
logarithmical dependence on the angle as for the one pair production.  In
the region $\theta > 0.2$ rad covered by the tracking system it is
1000 times smaller than the cross section for one pair. In principle,
one can install a fine segmented calorimeter at the angles down to 20
mrad, where backgrounds are acceptable, and detect the process
$\EPEM\EPEM$ with the cross section about 1 pb at $2E_0=500$ GeV with
a rather good energy resolution. Nevertheless, it can not compete with
the process of one pair lepton production which has more than one
order higher counting rate, better resolution on invariant masses and
much lower systematics.
\vspace{-5mm}
\section{Measurement of \GE\ luminosity}
\vspace{-5mm} Here two QED processes are of interest: $\GE\to\GE$ and
$\GE\to\;$e$^-\EPEM$. The cross section for the first process is
proportional to $\alpha^2/W^2_{\GE}\,$, the second one is of the higher
order on $\alpha$: $\alpha^3/(\WGE \theta)^2$, however at small angles
its cross section is even larger.

 \vspace{-0.2cm}

\begin{figure}[!hbt]
\hspace*{0cm} \vspace*{-0.0cm}  
\begin{minipage}[b]{0.45\linewidth}
\centering
\hspace*{-0.5cm} \epsfig{file=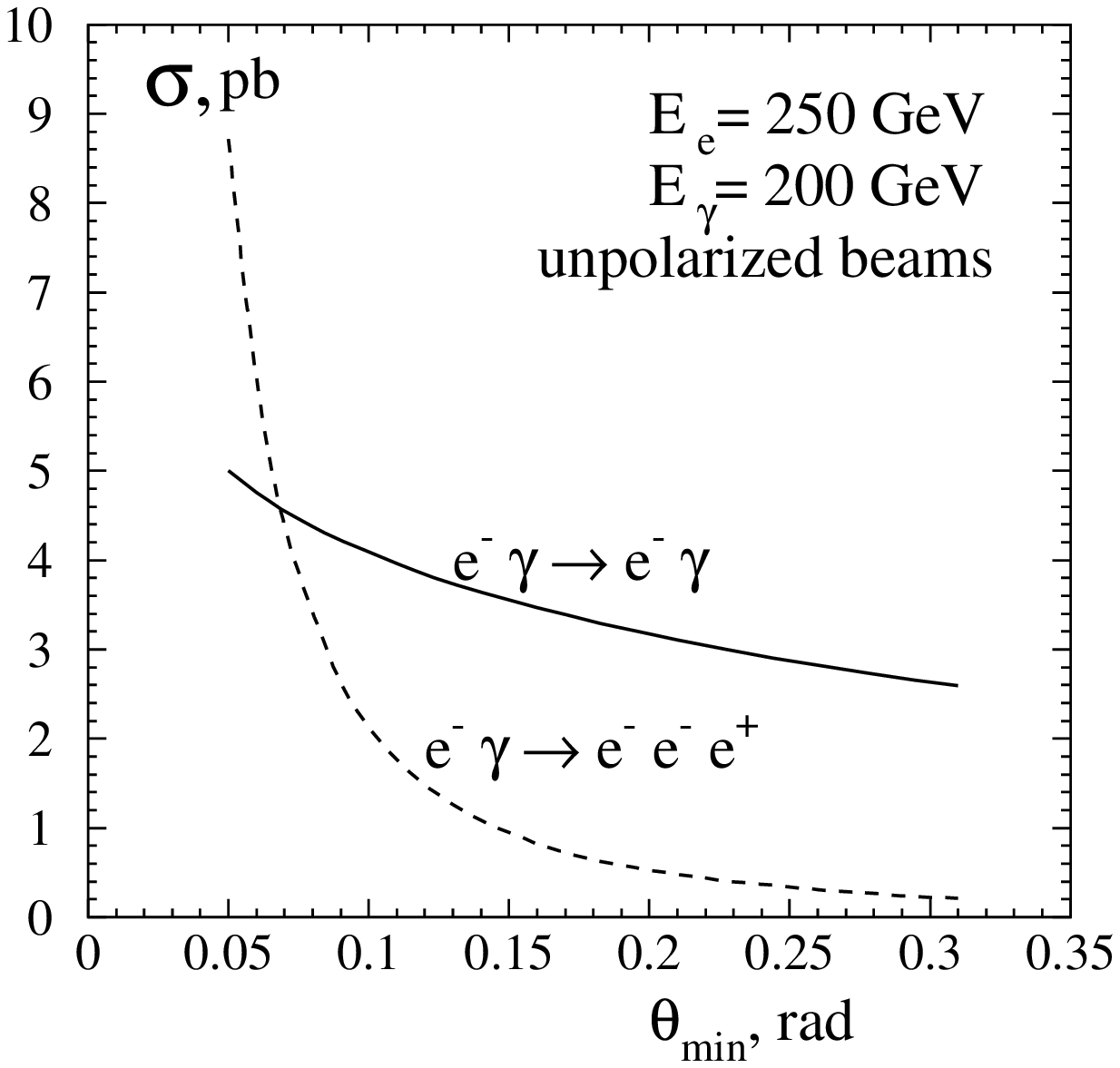,width=6.5cm,angle=0}
\vspace*{-0.6cm}
\caption{ \label{lge} The cross section of  processes for \GE\ luminosity
  measurement as a function of the minimum detection angle.}
\end{minipage}
\hspace*{1.cm} 
\begin{minipage}[b]{0.45\linewidth}
\centering
\vspace*{-0.0cm} 
\hspace*{-0.7cm} \epsfig{file=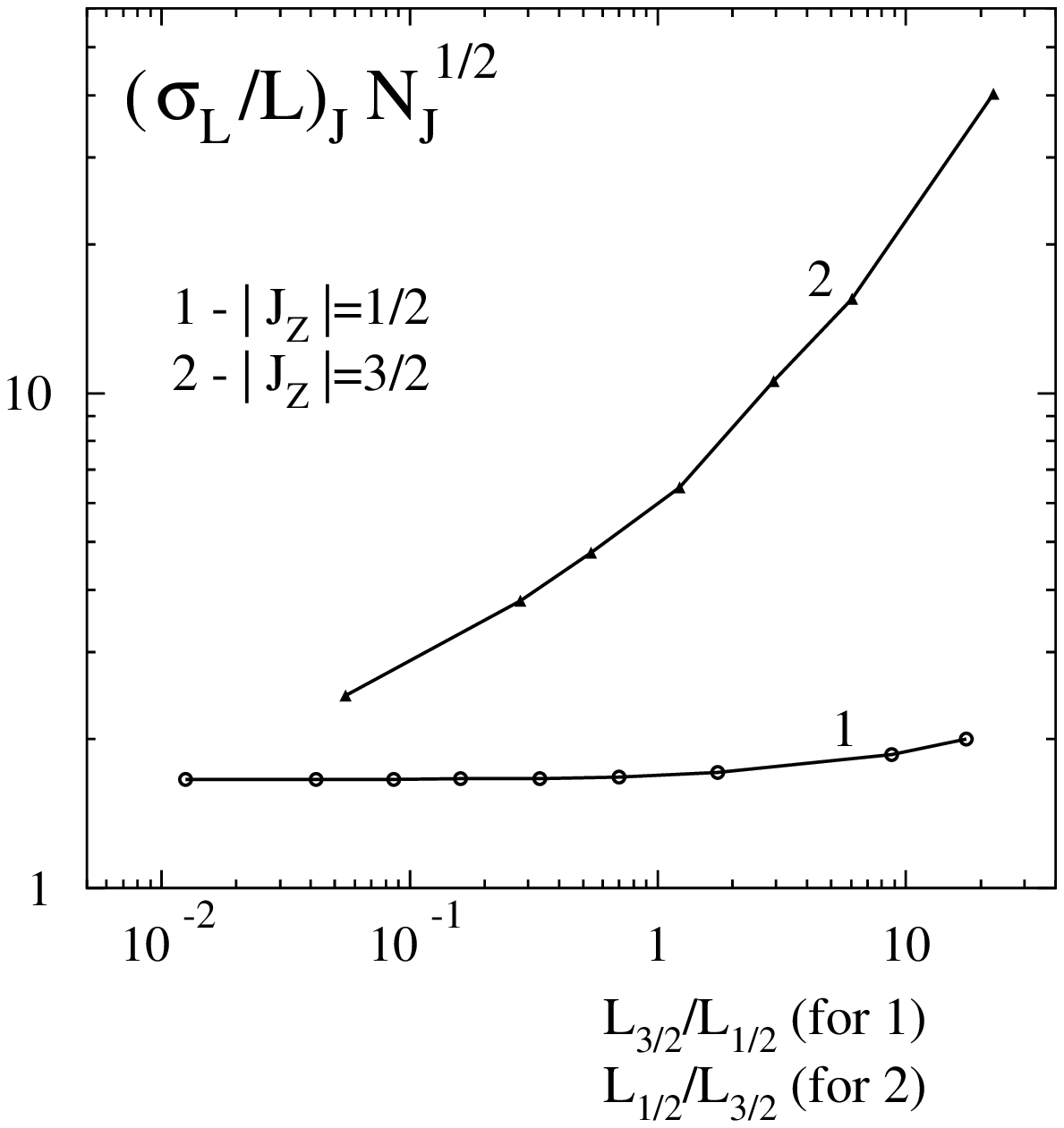,width=6.5cm,angle=0} 
\vspace*{-0.4cm}
\caption{\label{sumbat}  Statistical accuracy of   $L_{1/2}$ and $L_{3/2}$  
as a function of the luminosities ratio. }
\vspace*{0.3cm}
\label{ris1}
\end{minipage}%
\end{figure}

{\boldmath \underline{$\gamma e \to \gamma e$}}.  Cross section in c.m.s. at $\theta
\gg 1/\gamma$
 \vspace{-0.2cm}
\begin{equation}
d\sigma = \frac{\pi\alpha^2}{2\WGE^2}\left[(1-2\lambda_e\lambda_{\gamma})
(1-\cos \theta_{\gamma}) + 
(1+2\lambda_e\lambda_{\gamma})\frac{4}{1-\cos \theta_{\gamma}}\right]
d(\cos \theta_{\gamma}),
\label{cge}
 \vspace{-0.2cm}
\end{equation}
where $\lambda_{\gamma}$ and $\lambda_{e}$ ($|\lambda_{e}|<1$) are the photon and electron
helicities, z-axis is
along the initial direction of the electron. The first term
corresponds to the case when helicities of the electron and photon have
opposite signs, i.e. $|J_z| = 3/2$, the second, dominant term, to
$|J_z| = 1/2$. The  cross section for unpolarized
beams is shown in Fig.~\ref{lge}.

\vspace{-3mm}

Additional problem in \GE\ collisions is connected with the fact that
the photon can come from two directions (if both electron beams
collide with laser photons). Luminosities for each direction should be
measured independently.  According to (\ref{cge}), the cross section
with $|J_z| = 3/2$ and $|J_z| = 1/2$ are comparable at
$\theta_{\gamma} \sim \pi$ and, it seems, that one can measure both
luminosities $L_{1/2}$ and $L_{3/2}$  fitting the
angular distributions. However, due to two possible photon directions
the effective cross section is $ \sigma = (\sigma(\theta)+\sigma(\pi -
\theta))/2$, and $\sigma_{1/2}$ dominates at all angles. We have
checked the statistical accuracy of measurement of $L_{1/2}$ and
$L_{3/2}$. Let the number of produced events in these polarization
states be $N_{1/2}$ and $N_{3/2}$. In the ideal case $\sigma_{Li}/L_i
= 1/\sqrt{N_i}$, in reality the accuracies depend on the ratio of
luminosities as shown in Fig.~\ref{sumbat}. One can see that $L_{3/2}$
can be measured with an accuracy close to $1/\sqrt{N_i}$ only when
$L_{1/2} \ll L_{3/2}$.  For the measurement of both $L_{1/2}$ and
$L_{3/2}$ with a high accuracy one should invert polarization of one
beam to opposite in the same way as for \GG\ collisions. For \GE\ 
luminosity spectra at TESLA(500) (see Fig.\ref{Ldist250}) and beams with
equal polarizations the number of events in the high energy peak for
$10^7$ sec is $N_{1/2} = 3\times 10^{5}$ and $N_{3/2} = 5\times
10^{3}$. The luminosities $L_{1/2}$ and $L_{3/2}$ are measured with
accuracies 0.3\% and 20\%, respectively.  After inversion of
polarization for one beam, distributions for $L_{1/2}$ and $L_{3/2}$
replace each other and new number of events are $N_{1/2} = 5\times
10^{4}$, $N_{3/2} = 3\times 10^{4}$ and the accuracies for $L_{1/2}$
and $L_{3/2}$ are 0.8\% and 1.8\%, respectively.  Thus, the measurement with
inverted polarization improves the accuracy for $L_{3/2}$ 
(that was before the inversion) from 20 \% to 0.8\%.  

\vspace*{-0.3cm}

{\boldmath \underline{$\gamma e \to e^-e^+e^-$}.}
The cross section of the Bethe-Heitler process for the case when all
three final particles have angles above $\theta_{\min}$ is shown in
Fig.~\ref{lge}.  As was mentioned before, at $\theta < 70$ mrad it is
larger than that for $\GE\to\GE$.  It is not small even in
the region above 0.15--0.2 rad where detectors measure precisely
parameters of all particles. The cross section of this process only
weakly depends on the polarization of initial particles. Tohether with
the process $\GE\to\GE$ it allows to measure $L_{3/2}$ with
sufficiently good accuracy without the inversion of beam polarizations
or can be used for the cross check.

\vspace{-0.8cm}

\vspace{-0.3cm}

{\boldmath \underline{$\gamma e \to eZ$}.}  For study of some
processes, such as $\GE\to W\nu$, the independent measurement of
$\lambda_e$ and $\lambda_{\gamma}$ is necessary. In the Compton process
at $\theta \gg 1/\gamma$ the terms containing such dependence are
neglegibly small (see \cite{TESLATDR}). For the measurement  one can
use the process $\GE\to eZ$ (may be there is better one?).  Its
cross section for $ \WGE\gg M_Z$ and $\WGE\ \sim M_Z$, respectively,
is \cite{renard}
$$
\sigma \propto \frac{(1+2\lambda_e\lambda_{\gamma})-
0.2(2\lambda_e+\lambda_{\gamma})}{1-\cos{\theta_Z}}
+\frac{(1-2\lambda_e\lambda_{\gamma})- 0.2(2\lambda_e-\lambda_{\gamma})}{4}
(1-\cos{\theta_Z})
$$
\begin{equation}
\sigma \propto \frac{(1-2\lambda_e\lambda_{\gamma})-
0.2(2\lambda_e-\lambda_{\gamma})}{1-\cos{\theta_Z}}.
\end{equation}
One can see that it is possible to extract the electron and photon
polarization separately assuming that their product is measured by
$\GE\ \to \GE$ with a high accuracy.
\vspace{-9mm}
\section{Discussion, conclusion}
\vspace{-5mm}
The \GG\ luminosity for circular and linear polarized beams can be
measured using the process $\GG\ \to l^+l^-$.  Since this cross section is
negligible for $J_z=0$, to measure of $L_0$ one has to invert the
polarization of one beam for part of the time. The processes $\GG\ \to
l^+l^-\gamma$ is sensitive to $J_z=0$ but has too small cross section.
The process $\GG\ \to l^+l^-l^+l^-$ has too small cross section in the
large angle part of the detector with good measurement of particle
momenta. Detection of this process in the region of small angles (above 20
mrad) may be useful.

\vspace{-3mm}

In \GE\ collisions the luminosity and product
$\lambda_e\lambda_{\gamma}$ can be measured with a high accuracy by
means of the process $\GE\to\GE$. In order to measure both helicity
combinations, inversion of the polarization for one of beams is necessary.
The Bethe-Heitler process $\GE\to\,$e$^-$\EPEM has also sufficiently
large cross section, it can provide independent check of the first
method and allows to improve the accuracy of $L_{3/2}$ without
inversion of beam polarizations.

\vspace{-1mm}

This work was supported in part by INTAS (00-00679) and RFBR (02-02-17884).

\end{document}